\input harvmac
\font\ensX=msbm10

\noblackbox
\def\tl{\widehat \l}

\def\chl{{\check \lambda}}
\def\chr{{\check r}}
\def\ha{{\widehat\alpha}}
\def\tree{{\rm tree}}
\def\CC{\hbox{\ensX C}}

\def\a{\alpha}
\def\b{\beta}
\def\g{\gamma}
\def\l{\lambda}
\def\lb{\bar\lambda}

\def\mm{{\bf m}}
\def\nm{{\bf nm}}

\def\cN{{\cal N}}

\def\p{\partial}
 

\lref\HoogeveenTU{
  J.~Hoogeveen and K.~Skenderis,
  ``BRST quantization of the pure spinor superstring,''
  JHEP {\bf 0711}, 081 (2007)
  [arXiv:0710.2598 [hep-th]].
}

\lref\BerkovitsNG{
  N.~Berkovits and C.~R.~Mafra,
  ``Equivalence of two-loop superstring amplitudes in the pure spinor and  RNS
  formalisms,''
  Phys.\ Rev.\ Lett.\  {\bf 96}, 011602 (2006)
  [arXiv:hep-th/0509234].
}
\lref\MafraJH{
  C.~R.~Mafra,
  ``Four-point one-loop amplitude computation in the pure spinor formalism,''
  JHEP {\bf 0601}, 075 (2006)
  [arXiv:hep-th/0512052].
}
\lref\BerkovitsVI{
  N.~Berkovits and N.~Nekrasov,
  ``Multiloop superstring amplitudes from non-minimal pure spinor formalism,''
  JHEP {\bf 0612}, 029 (2006)
  [arXiv:hep-th/0609012].
}
\lref\BerkovitsNEW{
 Y.~Aisaka and N.~Berkovits,
  ``Pure Spinor Vertex Operators in Siegel Gauge and Loop Amplitude Regularization''
  arXiv: arXiv:0903.3443
}
\lref\BerkovitsPX{
  N.~Berkovits,
  ``Multiloop amplitudes and vanishing theorems using the pure spinor
  formalism for the superstring,''
  JHEP {\bf 0409}, 047 (2004)
  [arXiv:hep-th/0406055].
}
\lref\BerkovitsBT{
  N.~Berkovits,
  ``Pure spinor formalism as an N = 2 topological string,''
  JHEP {\bf 0510}, 089 (2005)
  [arXiv:hep-th/0509120].
}
\lref\BerkovitsVC{
  N.~Berkovits,
  ``New higher-derivative R**4 theorems,''
  Phys.\ Rev.\ Lett.\  {\bf 98}, 211601 (2007)
  [arXiv:hep-th/0609006].
}
\lref\BerkovitsZK{
  N.~Berkovits,
  ``ICTP lectures on covariant quantization of the superstring,''
  arXiv:hep-th/0209059.
}
\lref\GreenGT{
M.~B.~Green, J.~G.~Russo and P.~Vanhove,
``Non-renormalisation conditions in type II string theory and maximal
supergravity,''
JHEP {\bf 0702}, 099 (2007)
[arXiv:hep-th/0610299]; 
}
\lref\GreenYU{
M.~B.~Green, J.~G.~Russo and P.~Vanhove,
``Ultraviolet properties of maximal supergravity,''
Phys.\ Rev.\ Lett.\  {\bf 98}, 131602 (2007)
[arXiv:hep-th/0611273].
}
\lref\GreenSW{
  M.~B.~Green, J.~H.~Schwarz and L.~Brink,
  ``N=4 Yang-Mills And N=8 Supergravity As Limits Of String Theories,''
  Nucl.\ Phys.\  B {\bf 198}, 474 (1982).
}
\lref\BernXJ{
Z.~Bern, J.~J.~Carrasco, D.~Forde, H.~Ita and H.~Johansson,
``Unexpected Cancellations in Gravity Theories,''
arXiv:0707.1035 [hep-th].
}
\lref\BjerrumBohrJI{
  N.~E.~J.~Bjerrum-Bohr and P.~Vanhove,
  ``Absence of Triangles in Maximal Supergravity Amplitudes,''
  JHEP {\bf 0810}, 006 (2008)
  [arXiv:0805.3682 [hep-th]].
}
\lref\BadgerRN{
  S.~Badger, N.~E.~J.~Bjerrum-Bohr and P.~Vanhove,
  ``Simplicity in the Structure of QED and Gravity Amplitudes,''
  arXiv:0811.3405 [hep-th].
}
\lref\ArkaniHamedGZ{
  N.~Arkani-Hamed, F.~Cachazo and J.~Kaplan,
  ``What is the Simplest Quantum Field Theory?,''
  arXiv:0808.1446 [hep-th].
}
\lref\NekrasovWG{
  N.~A.~Nekrasov,
  ``Lectures on curved beta-gamma systems, pure spinors, and anomalies,''
  arXiv:hep-th/0511008.
}
\lref\GrassiNZ{
  P.~A.~Grassi, G.~Policastro and P.~van Nieuwenhuizen,
  ``Superstrings and WZNW models,''
  arXiv:hep-th/0402122.
}
\lref\OdaAK{
  I.~Oda and M.~Tonin,
  ``Y-formalism and $b$ ghost in the Non-minimal Pure Spinor Formalism of
  Superstrings,''
  Nucl.\ Phys.\  B {\bf 779}, 63 (2007)
  [arXiv:0704.1219 [hep-th]].
}
\lref\AnguelovaPG{
  L.~Anguelova, P.~A.~Grassi and P.~Vanhove,
  ``Covariant one-loop amplitudes in D = 11,''
  Nucl.\ Phys.\  B {\bf 702}, 269 (2004)
  [arXiv:hep-th/0408171].
}
\lref\Bernhh{
Z.~Bern, J.~J.~Carrasco, L.~J.~Dixon, H.~Johansson, D.~A.~Kosower and R.~Roiban,
``Three-Loop Superfiniteness of N=8 Supergravity,''
Phys.\ Rev.\ Lett.\  {\bf 98}, 161303 (2007)
[arXiv:hep-th/0702112].
%
Z.~Bern, L.~J.~Dixon and R.~Roiban,
``Is N = 8 supergravity ultraviolet finite?,''
Phys.\ Lett.\  B {\bf 644}, 265 (2007)
[arXiv:hep-th/0611086].
  Z.~Bern, J.~J.~M.~Carrasco, L.~J.~Dixon, H.~Johansson and R.~Roiban,
  ``Manifest Ultraviolet Behavior for the Three-Loop Four-Point Amplitude of
  N=8 Supergravity,''
  arXiv:0808.4112 [hep-th].
  Z.~Bern, J.~J.~M.~Carrasco and H.~Johansson,
  ``Progress on Ultraviolet Finiteness of Supergravity,''
  arXiv:0902.3765 [hep-th].
}
\lref\GrassiTV{
  P.~A.~Grassi and G.~Policastro,
  ``Super-Chern-Simons theory as superstring theory,''
  arXiv:hep-th/0412272.
}

\lref\Atick{J.~J.~Atick, J.~M.~Rabin and A.~Sen,
  ``An Ambiguity In Fermionic String Perturbation Theory,''
  Nucl.\ Phys.\  B {\bf 299}, 279 (1988).
J.~J.~Atick, G.~W.~Moore and A.~Sen,
  ``Some Global Issues In String Perturbation Theory,''
  Nucl.\ Phys.\  B {\bf 308}, 1 (1988).
  J.~J.~Atick, G.~W.~Moore and A.~Sen,
  ``Catoptric Tadpoles,''
  Nucl.\ Phys.\  B {\bf 307}, 221 (1988).
}
\lref\Martinec{
  E.~J.~Martinec,
  ``Nonrenormalization Theorems And Fermionic String Finiteness,''
  Phys.\ Lett.\  B {\bf 171}, 189 (1986).
}
\lref\lechtenfeld{
O.~Lechtenfeld,
  ``Factorization And Modular Invariance Of Multiloop Superstring Amplitudes In
  The Unitary Gauge,''
  Nucl.~Phys.~B~{\bf 338}, 403~(1990).
O.~Lechtenfeld and A.~Parkes,
  ``On Covariant Multiloop Superstring Amplitudes,''
  Nucl. Phys. B {\bf 332}, 39 (1990).
 O.~Lechtenfeld,
  ``On The Finiteness Of The Superstring,''
  Nucl.Phys.B {\bf 322}, 82~(1989).
}
\lref\GrassiIH{
  P.~A.~Grassi and L.~Tamassia,
  ``Vertex operators for closed superstrings,''
  JHEP {\bf 0407}, 071 (2004)
  [arXiv:hep-th/0405072].
}
\lref\SiegelSV{
  W.~Siegel and H.~d.~Feng,
  ``Gauge-covariant vertex operators,''
  Nucl.\ Phys.\  B {\bf 683}, 168 (2004)
  [arXiv:hep-th/0310070].
}
\lref\ZwiebachIE{
  B.~Zwiebach,
  ``Closed string field theory: Quantum action and the B-V master equation,''
  Nucl.\ Phys.\  B {\bf 390}, 33 (1993)
  [arXiv:hep-th/9206084].
}
\lref\BerkovitsHG{
  N.~Berkovits,
  ``Finiteness and unitarity of Lorentz covariant Green-Schwarz superstring
  amplitudes,''
  Nucl.\ Phys.\  B {\bf 408}, 43 (1993)
  [arXiv:hep-th/9303122].
}
\lref\MafraWQ{
  C.~R.~Mafra,
  ``Superstring Scattering Amplitudes with the Pure Spinor Formalism,''
  arXiv:0902.1552 [hep-th].
}
\lref\BerkovitsBK{
  N.~Berkovits and C.~R.~Mafra,
  ``Some superstring amplitude computations with the non-minimal pure spinor
  formalism,''
  JHEP {\bf 0611}, 079 (2006)
  [arXiv:hep-th/0607187].
}

\Title{\vbox{\hbox{DISTA-2008, IPHT-T-08/019, IHES-P-09-13, NSF-KITP-09-31}}}
{\vbox{ 
\centerline{Higher-loop amplitudes}
\centerline{in the non-minimal pure spinor formalism}}} 
\medskip\centerline{Pietro Antonio Grassi${}^{a,b,e}$
and  
Pierre Vanhove${}^{c,d,e}$}
\medskip 
\centerline{$^{a}$ \sl DISTA, Universit\`a del Piemonte Orientale, 
 via Bellini 25/g, 15100 Alessandria, ITALY}
\centerline{$^{b}$ \sl INFN, gruppo collegato sezione di Torino, ITALY}
\centerline{${}^c$  \sl IHES,  Le  Bois-Marie, 35  route de  Chartres,
F-91440 Bures-sur-Yvette, France}
\centerline{${}^d$  \sl CEA,  DSM, Institut  de  Physique Th\'eorique,
IPhT,}
\centerline{\sl CNRS,  MPPU, URA2306, Saclay,  F-91191 Gif-sur-Yvette,
France}
\centerline{${}^e$ \sl Kavli Institute for Theoretical Physics,} 
\centerline{\sl University of California at Santa Barbara, CA 93106-4030, USA}
\bigskip\bigskip
 \centerline{\tt email: pgrassi@cern.ch,\    pierre.vanhove@cea.fr}
\medskip
\vskip  .5cm
\centerline{\bf Abstract}

We analyze  the properties of  the non-minimal pure  spinor formalism.
We show  that Siegel  gauge on massless  vertex operators  implies the
primary field  constraint and  the level-matching condition  in closed
string  theory  by   reconstructing  the  integrated  vertex  operator
representation from the unintegrated ones. The pure spinor integration
in the non-minimal  formalism needs a regularisation.  To  this end we
introduce  a new  regulator for  the pure  spinor integration  and an
extension  of  the  regulator  to  allow for  the  saturation  of  the
fermionic $d$-zero  modes to all orders in  perturbation.  We conclude
with  a preliminary analysis  of the  properties of  the four-graviton
amplitude to all genus order.

\Date{\the\day/\the\month/\the\year}

\newsec{Introduction}

The   pure   spinor   formulation   of  perturbative   string   theory
\refs{\BerkovitsPX,\BerkovitsBT} has proved to  be a powerful tool for
implementing   the   role  of   maximally   extended  ${\cal   N}=8$
supersymmetries in various amplitude computations.  Because this formalism makes use of a constrained ghost variable it allows
to construct  superspace invariants  over fraction of  superspace coordinates that
are difficult to construct  in conventional superspace approaches.  In
an extended formulation of the  pure spinor formalism, Berkovits was able
to avoid the complications associated  with the picture changing operators of
the  original  multiloop prescription~\refs{\BerkovitsBT,\BerkovitsVI}
and    to    obtain    a    new   class    of    partial    superspace
integrals~\refs{\BerkovitsVC} giving  the leading contribution  to the
low-energy limit of the four-gravitons amplitude at genus order $g\leq
6$ 
\eqn\eFterm{
 F_g=  \int    d^{16}    \theta    d^{16}\bar\theta
\theta^{12-2g}\bar\theta^{12-2g}                                     \,
(W_{\alpha\beta})^4\sim \partial^{2g} R^4+ {\rm susy~completion}
}
Where    $W_{\alpha\beta}$    is     the    Ramond-Ramond    spin    1
superfield~\refs{\BerkovitsZK,\BerkovitsVC}. 
The fact  that these quantities  give the leading contribution  to the
low-energy  limit  of the  four-graviton  amplitude,  up to  genus-six
order, confirms the non-renormalisation conditions
for the $\partial^{2g}R^4$ contributions with $g\leq6$ to the ten dimensional
low-energy effective  action for type~IIA and  type~IIB string derived
from  string   dualities  in  \refs{\GreenGT}.

Since these superspace integrals arise from the zero mode saturation they
 give a direct indication of the leading ultra-violet divergence structure
of the field theory four-graviton amplitude in $\cN=8$ supergravity. A
four-graviton  amplitude with  the leading  low-energy limit  given by
$F_g$ in~\eFterm\ has the following dimensions
by 
\eqn\eAmpC{
[{\cal  A}^g_4] =  [\partial^{2g}R^4]\, {\rm  mass}^{(D-4)g-6}\, \qquad
g\leq 6
}
where $[\cdots]$ gives the mass dimension. We used
that  a   $g$-loop  gravity  amplitude  has   mass  dimension  $[{\cal
A}^g_4]={\rm   mass}^{(D-2)g+2}$,     that   $[\partial]=$mass  and
$[R^4]={\rm mass}^8$.
 It is
remarkable  that the  explicit four-graviton  amplitudes  performed in
field  theory  up  to and  included  three  loop  order in~\refs{\GreenSW,\Bernhh} can  be
presented in a form that has the manifest ultra-violet
behaviour given by~\eAmpC.  This formula indicates that the $g$-loop four-graviton
amplitude in~\eAmpC\  develops ultra-violet divergences from
\eqn\eDc{
D\geq D_c=4+{6\over g}; \qquad g\leq 6\ .
}
When $g=6$ the integration in~\eFterm\ is over all the full superspace
(all the 32 $\theta$
variables)   and  supersymmetric  protection  is
exhausted.   But   at  precisely   this  order  the   amplitudes  are
ill-defined because of singularities  in the integration over the pure
spinor ghosts~\refs{\BerkovitsVI,\BerkovitsVC} and no firm conclusions
could be  drawn about the  structure of the amplitude  at higher-genus
order. 
In this work we discuss an alternative modification of the non-minimal
pure spinor formalism  leading to well defined amplitude  at any genus
order.  
 A regularisation of the singularities from the tip of the cone has been given
in~\refs{\BerkovitsVI} but  the resulting formulation makes very difficult
to  extract   information about  the   structure  of  the
higher-loop amplitudes. In order  to understand the systematics of the
higher-loop multigraviton amplitudes we introduce  an
alternative  regulator.  With this  regulator  we  give a  preliminary
analysis of the structure of
the four-graviton amplitude at higher-genus.
We hope that this  analysis is a step toward understanding the
systematics of     ${\cal
N}=8$ supergravity amplitudes and the role of the surprising simplifications
occurring       the       structure       of       the       higher-loop
amplitudes~\refs{\GreenGT,\Bernhh,\GreenYU,\BernXJ,\BjerrumBohrJI,\BadgerRN,\ArkaniHamedGZ}.

 In  section~2  we  review  the  basics of  the  minimal  pure  spinor
 formalism  and  its  relation   to  the  non-minimal  formalism.   In
 section~3,  we   discuss  the   massless  vertex  operators   in  the
 non-minimal formalism.  We derive the relation between the integrated
 and  unintegrated representation  of the  vertex operators.   Using a
 Siegel  gauge we  derive  the physical  state  condition on  massless
 vertex operators, and the level-matching condition in the case of the
 closed string. Because of the  dependence of the $b_\nm$-ghost on the
 non-minimal  sector  the  change  of  representation  of  the  vertex
 operator and  the Siegel gauge are  only obtained up  to Q-exact term
 depending  on the non-minimal  sector.  A  different analysis  of the
 Siegel  gauge  condition  on  vertex operators  appeared  the  recent
 preprint~\refs{\BerkovitsNEW}.  In section~4 we analyze the
origin   of  divergences   in   the  pure   spinor  integration.   The
singularities in  the pure  spinor integration are  taken care  by the
introducing of  a new
regulator  strongly dumped  at the  tip  of the  cone.
 We show that in order to be
 able  to  saturate  the  fermionic   zero  modes  to  all  orders  in
 perturbation---and avoid that the amplitudes are  vanishing
 after some genus order which would be incompatible with unitarity---one needs to consider an extension of the regulator with
 more $d$-zero mode contributions. 
In  our  scheme the  non-minimal  $b_\nm$-ghost  is  not modified  and
applies to any genus order and any number of punctures.  In section~5 we turn to multiloop
 amplitudes  and  give  the  form  of the  integrand  of  the  leading
 low-energy contribution  to the multiloop  four-graviton amplitude at
 all  genus   order.   We  conclude  by  showing   that  the  massless
 $N<4$-point amplitudes are vanishing  to all order in the non-minimal
 pure   spinor   formalism.   This   implies   finiteness  of   string
 perturbation  in  the  absence  of unphysical  singularities  in  the
 interior of the moduli space.

\newsec{Pure  spinor  measure  of   integration  in  the  minimal  and
non-minimal  formalism} 

The action  for type  II superstring in  the pure spinor  formalism in
flat ten-dimensional  space 
is given by~\refs{\BerkovitsPX}
\eqn\ePsaction{
S= \int d^2z \, \left({1\over2\pi\alpha'}\partial x^m\bar\partial x_m + p_\alpha \bar\theta^\alpha+ \widehat p_{\ha} \partial \widehat\theta^{\ha} + w_\alpha\bar\partial\lambda^\alpha +\widehat w_{\ha}\partial\tl^{\ha} \right)
} 
The  matter fields  are organized into ten bosonic fields  of conformal
weight   zero  $x^m$   with  $m=0,\dots,9$   and  two sets of fermionic  fields
$(p_\alpha,\theta^\alpha)$       and       $(\widetilde       p_{\ha},
\widetilde\theta^{\ha})$  of conformal  weight one  and zero  with  $\a$ in
${\bf  16}$ and  $\ha$ in  ${\bf 16}$  or $\bar{\bf  16}$  of $SO(16)$
depending if  one treats  the type~IIA or  type~IIB string.  In the following we will only mention the left-moving sector, but there are identical contributions from the right-moving sector. The pure
spinor ghost $\lambda^\alpha$ of conformal weight zero is constrained by 

\eqn\ePS{\l \g^m \l =0}
where $(\g^m)_{\a\b}$ are the $16\times16$ gamma matrices of $SO(10)$.
The pure spinor space  defined by the constraint~\ePS\ is
  the non-compact conical space defined by a $\CC^\star$ bundle over $SO(10)/U(5)$.  The  scale of  the pure  spinor varies between  0 and
$\infty$.

The constraint leaves 11 independent components for the pure spinor $\l^\a$  and implies that the conjugated pure  spinors $w_{\a}$ of conformal weight one  has the following $\Lambda$-gauge invariance  $\delta_\Lambda w_{\a}  =
\Lambda_{   m}   (\g^m    \l)_{\a}$   with   $\Lambda_{m}$   a   gauge
parameter. 
The physical quantities are described as the cohomology of the pure spinor BRST charge
\eqn\eQB{
Q_\mm=\oint \lambda^\a \, d_\a
}
where
$
d_\a= p_a - {1\over2} \, (\g^m\theta)_\a \partial x_m -{1\over8}(\theta\g_m\partial\theta)(\g^m\theta)_\a
$
is the Green-Schwarz constraint, which satisfies the OPE
$
d_\a(z)\, d_\b(0) \sim- (\g^m)_{\a\b} \, \Pi_m/z
$
where  $\Pi_m=  \partial x_m  +(\theta\g_m  \partial\theta)/2$ is  the
supersymmetric momentum. Analogously for the right-moving sector.

In the case of the minimal formalism~\refs{\BerkovitsPX} at genus $g$ order,
the $11$ zero modes of the pure spinor ghost $\lambda^\a$ and $11g$ zero modes for the conjugated ghost $w_\a$ are saturated by the insertions of delta-functions $\delta(\l^\a)$ and $\delta(w_\a)$. 
The BRST-invariant and $\Lambda$-gauge invariant version of these delta-functions  is given  by 
the  picture lowering  $Y_C$  and the picture   raising $Z_B$  operators

\eqn\pco{
Y_C = C_\a \theta^\a\, \delta(C_\a \l^\a)\,, ~~~~~
Z_B = \Big[ Q_\mm, \Theta( [wB\l]) \Big]=(dB\l)\,\delta(wB\l)\ ,
}
where $\Theta$ is the Heaviside step-function, and we have made use of the
following notation 
\eqn\ewl{
[wB\l]\equiv \, :w_\a \, B^\a{}_\b\, \l^\b:\, = B\, J+
{1\over2!}\, B_{mn}\, N^{mn}
}
where the gauge-fixing parameters are the constant spinor $C_\a$, and the
46  constants  $B$ and $B_{mn}$.  We have  as well introduced
the currents 
\eqn\eNJ{
J = :w_\a \l^\a: \qquad N^{mn} =:w\g^{mn} \l:
}
are conformal weight one $\Lambda$-gauge invariant quantities. 

The integration over the bosonic moduli is taken care
by the picture raised conformal weight two $b_\mm$-ghost which satisfies $[b_\mm , Q_\mm] = Z_B\,
T_\mm$ where $T_\mm$ is the minimal formalism stress energy tensor.
This field  is  integrated over the Riemann surface $\Sigma_g$ with the help of the Beltrami differentials $
(\mu|b_\mm) \equiv \int_{\Sigma} d^2z \, \mu^z{}_{\bar z} \, b_{\mm\,zz}
$
and the prescription for a genus-$g$ amplitude, with $g\geq2$, in type
IIA/IIB  string  theory  is   given  by  \refs{\BerkovitsPX}  (see  as
well~\refs{\HoogeveenTU}  for an  alternative derivation  of  the pure
spinor measures)
\eqn\nmB{\eqalign{
{\cal A}^g_N &=\int d^{3g-3}\tau\,\left\langle \Big| \prod_{i=1}^{3g - 3} (\mu_i| b_\mm) \prod_{j=3g -2}^{11g} Z_{B_j} \prod_{k=1}^{11}Y_{C,k} \Big|^2 \, \prod_{i=1}^N V_i\right\rangle 
}} 
$V_i$ are  the integrated vertex  operators and $\langle\cdots\rangle$
represents  the  functional integration  over  the world-sheet  fields
$[x^m,p_\a,\theta^a,\l^\a,w_\a]$ is defined by 
\eqn\eIntM{
\langle \cdots \rangle=
\int d^{10}x d^{16}\theta \prod_{I=1}^g d^{16}d^I \int [d\l] \prod_{I=1}^g\int [dw^I]\cdots e^{-S_{ps}}
}
 At tree-level there is no
$w$-zero  mode and  the amplitude  is given  by 3  unintegrated vertex
operators  and  no insertions  of  $b_\mm$-ghost  of picture  changing
operators $Z_B$. At genus
one there are  $11$ $w$-zero mode to be integrated  over, there is one
insertion   of  the   $b_\mm$-ghost   and  one   vertex  operator   is
unintegrated. The  insertion of  the picture changing  operators $Y_C$
cuts off  the large  value of the  pure spinor $\l_\a$  localizing the
integration measure in a point.

The pure spinor measure of integration is defined as

\eqn\edlBerk{
[d\l]=   (\epsilon  {\cal  T}^{-1})^{\a\b\g}_{k_1\cdots
k_{11}} d\l^{k_1}\cdots d\l^{k_{11}}\,\p_{\l^\a}\p_{\l^\b}\p_{\l^\g}
}
where we have introduced the following tensor totally antisymmetric on the $k_i$
indices and fully symmetric
$\gamma$-traceless on the $\a\b\g$ indices~\refs{\BerkovitsBT}
\eqn\eeT{(\epsilon{\cal                  T})^{k_1\cdots
k_{11}}_{\a\b\g}=\epsilon_{16}^{k_1\cdots k_{11}r_1\cdots r_5}\, (\g^m)_{((\a
|r_1|}\, (\g^n)_{\b  |r_2|}\, (\g^p)_{\g ))r_3}\, (\g_{mnp})_{r_4r_5}\
.
}
Such a definition of the measure of integration using derivatives is
natural   from    the  supergeometry    point   of   view    as   shown
in~\refs{\GrassiTV}.
This measure  satisfies the requirement  that the overlap  between the
vacuum  $|0\rangle$  and  the  highest  state  in  the  zero  momentum
cohomology                                                  $|C\rangle=
(\l\g^m\theta)(\l\g^n\theta)(\l\g^p\theta)(\theta\g_{mnp}\theta)$ 
is a constant
\eqn\eTT{
\langle 0|C\rangle=\left\langle\prod_{i=1}^{11}
\theta^{\a_i}\delta(\l^{\a_i})
(\l\g^m\theta)(\l\g^n\theta)(\l\g^p\theta)(\theta\g_{mnp}\theta)\right\rangle=1\ .
}
This     gives     the      rules     for     computing     tree-level
amplitudes~\refs{\BerkovitsPX}.  We  will  return  to  this  computation  in
section~5  when analyzing the effect of the regulator on the
non-minimal formalism amplitude prescription.

This minimal formalism with only one set of pure spinor ghost, only
a picture raised version of the $b$-ghost can be constructed which make
the   analysis   of   the   multiloop   amplitude   difficult   beyond
two-loops. As  well in  this formalism the  integration over  the pure
spinor variables has to be done over
patches  of  the  pure spinor  space  and  one  needs to  analyze  the
\v{C}ech-cohomology       on       this       space       for       global
properties~\refs{\NekrasovWG}.  As well because of the presence of picture
changing operators the amplitudes are Lorentz and
supersymmetric invariant up to boundary term.

The delta-function insertions provided by the picture changing operators in~\pco\ can be exponentiated
by          introducing          extra          new          variables
\refs{\AnguelovaPG,\GrassiNZ,\BerkovitsBT}. 
Let start by considering the case a single fermionic variable $\theta$ whose BRST transformation is $Q \theta = \lambda$ and 
then by  adding a  new doublet $r$  and $\bar\l$ and  their conjugated
ghost $\bar  w$ and $s$  so that $[\bar  w, \bar\l] = 1$  and $\{r,s\}
=1$. In order that physical observables do not depend on these new variables,\foot{The
physical  vertex operators  do not  depend on  the  non-minimal sector
because    the   non-minimal   ghost    number   $\bar    J=   \lb\bar
w-sr=[Q_\nm,s_\a\lb^\a]$,  and  as well  $[Q_\nm,\bar  J]=0$. And  the
physical states are eigenvalues  of the non-minimal ghost number $\bar
J\,\Psi=n\, \Psi$. Since $\bar J$ is  Q-exact all
states   with   non-zero   non-minimal   ghost  charge   are   Q-exact
$\Psi=[Q_\nm, s\lb\,\Psi]/n$. Therefore the physical states are in the
zero oscillator sector with $n=0$. This is the so-called quartet mechanism.
} we introduce a new   nilpotent BRST operator $\Delta = \oint \bar w r$
 so that $(r,s;\bar\l,\bar w)$ is a topological quartet under the  total BRST-charge $Q+\Delta$.
We can now express  the delta-function insertions as follows 
\eqn\gfA{
\theta \delta(\l)\, d \delta(w) = 
\int [dr] [d\bar\l] [ds][d\bar w] \, {\cal N}
}
where
\eqn\cNa{
{\cal N}=e^{-\l\lb  - r \theta - w\bar w - s d }\
.
}
The exponent can be rewritten as $\bar\l \l + r \theta + w\bar
w + s d = [Q, \Psi]$ with the gauge fermion
\eqn\ePsiS{
\Psi=\bar\l \theta +  s w\ .
}
The form of the exponent as BRST-exact quantity ensures that the  amplitudes do not depend upon the extract form of the gauge fermion $\Psi$
 unless some singularities in the amplitude forbid the decoupling of BRST-exact quantities.

This procedure can be seen as a motivation for the introduction of the
non-minimal  ghosts  by Berkovits  in~\refs{\BerkovitsBT}  for defining  the
non-minimal pure spinor formalism. He introduced 
the complex conjugate extra ghosts $\bar\lambda_\a$ and $r_\a$ satisfy the relations

\eqn\connonmin{
\bar\l \g^m \bar \l = 0\,, ~~~~~
\bar\l \g^m r =0
}
 In this case and the conjugated variables transform under the gauge symmetry
$\delta_{\Lambda+L} \bar  w^\a =  \Lambda_m (\g^m \bar \l)^\a + L_m (\g^m r)^\a$
and  $\delta_L s^\a  = L_m  (\g^m \bar  \l)^\a$ where  $\Lambda_m$ and
$L_m$ are the gauge parameters. Therefore the conjugated ghost $\bar w^\a$ and
$s^\a$ can only appear through the conformal weight one $\Lambda$- and
$L$-gauge invariant quantities 
\eqn\eNN{\eqalign{
\bar N_{mn} &=  \bar w\g_{mn}\bar\l -
s\g_{mn} r; \qquad
 \bar J= \bar w\bar \l-s r\cr
S^{mn}&=s\gamma^{mn}\lb;\qquad S=s\lb\ .}}
  The non-minimal BRST-charge is 
\eqn\eQmn{
Q_{\nm} =\oint \lambda^\a d_\a +\oint \bar w^\a\, r_\a\, .
}

\newsec{Vertex operators in the non-minimal}

The  physical  state  vertex   operators  are  in  the  cohomology  of
$Q_{\nm}$ defined in~\eQmn.  For the massless sector of the type~II superstring the
vertex  operators come  into  the  integrated  and
the unintegrated representations
\eqn\eVV{
 V=\int  d^2z |V_{open}|^2\ e^{ik\cdot  X}; \qquad  U=|U_{open}|^2\,
 e^{ik\cdot X}
}
where $U_{open}=\l^\a A_\a$ and  
\eqn\eVop{
V_{open} =\partial \theta^\alpha A_\alpha +\Pi^m A_m
+ d_\alpha W^\alpha +{1\over2} N^{mn} F_{mn}
}
where  $A_\a$,  $A_m$,  $W^\a$   and  $F_{mn}$  are  the  $\cN=1$  $D=10$
super-Yang-Mills superfields
\eqn\eSuperfield{
A_\a(x,\theta)={1\over2}\,    (\g^m\theta)_\a   \,    a_m   +{1\over3}
(\chi\g_m\theta)\,(\g^m\theta)_\a -{1\over32} \, f_{mn} (\g_p\theta)_\a(\theta\g^{mnp}\theta)+\cdots
}
and 
\eqn\eEOM{\eqalign{
(\g^m)_{\a\b} A_m& = D_\a A_\b+D_\b A_\a\cr
(\g_m)_{\a\b} W^\b&= D_\a A_m-\p_m A_\a\cr
D_\a W^\b&={1\over 4}(\g^{mn})_\a{}^\b F_{mn}\cr
}}
 Acting with $Q_\nm$  on $V_{open}$ the computation is  the same as in
 the minimal formalism leading to
\eqn\eQV{
[Q_\nm, V_{open} ] =\partial_\sigma(U_{open}) +{\rm e.o.m.}
}
where e.o.m. are the $\cN=1$ $D=10$ super-Yang-Mills equations-of-motion given in~\eEOM.
The vertex operator $U_{open}$ satisfies 
$[Q_\nm, U_{open} ]  = 0$. Notice that, since  $V_{open}$ and $U_{open}$
are independent of the non-minimal fields only the 
minimal part of the BRST  charge acts on the vertex operator.

Because 
$\{Q_\nm, b_\nm\} = T_\nm$, one can use the $b_\nm$-ghost to construct
the   integrated  vertex   operator  from   the   unintegrated  vertex
operator. If we denote $b_{-1} = \int 
d\sigma\, b_\nm$, we have that 
$\{Q_\nm, b_{-1} \} =\int d\sigma \,T_\nm =\partial_\sigma$. So, acting with $b_{-1}$  on $U_{open}$ we can derive the integrated 
vertex operator $V_{open}$.

The non-minimal $b_{\nm}$-ghost takes the form \refs{\BerkovitsBT,\BerkovitsVI}
\eqn\nnA{
b_{\nm}= s \p \bar\l +{1\over4}\chl_\a \, {\bf b}^\a}
where we have introduced the notations

\eqn\eChLr{
 \chl_\a = {\bar\l_\a\over(\l\cdot\lb)};\qquad
 \chr_\a =  {r_\a\over(\l\cdot\lb)}}
and 
\eqn\eBB{
{\bf b}^\a \equiv G^\a+ \chr_\b H^{\a\b}+ \chr_\b\chr_\gamma
K^{\a\b\gamma}+\chr_\b\chr_\gamma\chr_\delta L^{\a\b\gamma\delta}}
and the operators
\eqn\eGHKL{\eqalign{
G^\a&\equiv2\Pi^m(\g_m d)^\a -N_{mn}(\g^{mn}\partial \theta)^\a-J\partial\theta^\a -{1\over2} \partial^2 \theta^\a\cr
H^{\a\b}&\equiv {1\over 192} (\g^{mnp})^{\a\b} \,\left((d\g_{mnp}d) +4! N_{mn}\Pi_p\right) \cr
K^{\a\b\g}&\equiv {1\over16} (\g_{mnp})^{[\a\b} (\g^md)^{\g]} N^{np}\cr 
L^{\a\b\g\delta}&= {1\over128} (\g_{mnp})^{[\a\b}(\g^{pqr})^{\g\delta]} N^{nm} N_{qr}  \ .
}}
It was shown in~\refs{\OdaAK} that the
non-minimal  $b_\nm$-ghost and  the $b_Y$-ghost  of  the $Y$-formalism
are related by
\eqn\ebY{
b_\nm = b_Y+ [Q_\nm, \Omega_v], \qquad b_Y={v_\a G^\a\over v\cdot \l}
}
where $v_\a$ is  a constant reference pure spinor  so that $v\g^m v=0$
and $v\cdot
\l\neq0$.  Here $\Omega_v$, which expression can be found
in~\refs{\OdaAK}, depends on the non-minimal sector and the
reference spinor $v_\a$.

We  want to  derive the integrated vertex operators $V_{open}$ by acting with $b_{\nm\,-1}$ on the unintegrated vertex
operators $U_{open} = \lambda^\a A_\a$. This amounts into taking the first order poles of the OPE
between the $b_\nm$-ghost and the vertex operator. 
For doing this computation we  will use the relation~\ebY\ and compute
the OPE between the Oda-Tonin $b_Y$-ghost with the vertex operator.

Using the ten-dimensional identity~\refs{\OdaAK}
\eqn\eFierz{
- {1 \over 8} (B\gamma^{mn}A)(\gamma_{mn}C)^\alpha - {1 \over 4}(B_\b A^\b) C^\a = 
B_\b A^\a C^\b - {1 \over 2} (\gamma^m B)^\a (A \gamma_m C)
}
where $A^\a, B_\gamma, C^\beta$ are three spinors of different chirality. 
It follows from the usual Fierz identities
and the OPEs 
\eqn\eNL{\eqalign{
N^{mn}(y)\,\l^\a(z)&\sim {1\over 2\,(y-z)}\, (\g^{mn}\l)^\a(0),\cr
J(y) \,\l^\a(z)&\sim      {1\over y-z}\, \l^\a(z)
}}
and  the  equations of motion  given in~\eEOM\ and the  Feynman gauge
condition $\p_m A^m =0$ we get 
\eqn\ebUT{\eqalign{
\oint_z b_Y(z)\,  (\l\cdot A)(0)&=  \p\theta^\a A_\a+ \Pi^m  A_m+ d_\a
W^\a+ {1\over2}F_{mn}N^{mn} + [Q_\nm, \widehat\Omega]
}}
where (all the pieces should be normal ordered)
\eqn\eOm{
\widehat\Omega=
 {( v \gamma^m A)\over 2(v\cdot \lambda)} \Pi_m - {( v \gamma^m d)\over (v\cdot \lambda)} A_m - {(v \gamma^m \partial_m A) \over (v\cdot \lambda)}
{\partial (v\cdot \lambda) \over (v\cdot \lambda)} + N^{mn} {(v \gamma_{mn} W)\over (v\cdot \lambda)} + {1\over 2} J {(v\cdot W)\over (v\cdot \lambda)}\,,
}
The  Q-exact part  in eq.~\ebUT\  contains all  the dependence  on the
auxiliary constant spinor $v$ and is needed as well for generating the
 $N_{mn} F^{mn}$ piece of the vertex operators.

This gives for the action of the full non-minimal $b_\nm$-ghost that
\eqn\eVopBis{
 \oint_z (b_\nm U_{open} + [Q_\nm, \widehat\Omega])=V_{open}
} 
where the Q-exact part assures  that $V_{open}$ does not depend on the
non-minimal sector.

 \subsec{Siegel Gauge for open string}

Within  the  pure spinor  formalism,  one  can  verify that  the  BRST
cohomology at ghost number one gives the superspace equations for $\cN=1$ 
$D=10$  the super-Yang-Mills. However,  these equations
are  not enough  to  impose  the primary field  constraint  on the  vertex
operator.  This situation is  well-known for  example in  String Field
theory where  the equations of  motion are manifestly  gauge invariant
(see  also~\refs{\SiegelSV}).   In  order  to   impose  the  primary field
condition, we impose the Siegel gauge condition.

For this we use the Oda-Tonin $b_Y$-field given in~\ebY. We define the 
zero mode of it as $b_{Y\,0} = \oint dz \, z \, b_{Y}$ and we act on the vertex operator 
$U_{open} = \lambda^\a A_\a(x,\theta)$. Computing the contributions of
the double poles yield
\eqn\actA{\eqalign{
b_{Y\,0} (U) &= {v_\a \lambda^\b\over v\cdot\lambda}  \, (\gamma^m)^{ \a\g} \partial_m D_\gamma 
\, A_\b\cr
& =  {1\over v\cdot\lambda} \, v_\a \lambda^\b 
(\g^m )^{\a\g} \Big( - D_\b \partial_m A_\g + (\g^p)_{\b\g} \partial_m A_p \Big)
\cr
&= - {1\over v\cdot\lambda}  \lambda^\b D_\b \Big( v_\a (\g^m )^{\a\g} \partial_m A_\g\Big) + 
{1\over v\cdot\lambda} v_\a\, \lambda^\b (\g^m \g^p)^\a_{~\b} \p_m A_p
\cr
&= - Q_\nm \left( { v_\a (\g^m)^{ \a\g} \partial_m A_\g \over v\cdot\lambda} \right) + 
{1\over v\cdot\lambda} \Big( v \g^m \g^p \lambda\Big) \p_m A_p
\cr
&= - Q_\nm \left( { v_\a (\g^m)^{ \a\g} \partial_m A_\g \over v\cdot\lambda} \right) + 
\p^m A_m + 
{1\over 2\, v\cdot\lambda} \Big( v \g^{mp} \lambda\Big) F_{mp}
}}
and finally, using again the equation of motion $4\,D_\a W^\b = (\g^{mn})_\a{}^\b F_{mn}$
we have 
\eqn\actB{
b_{Y\,0} (U) = \p^m A_m - Q_\nm\left(  { v_\a (\g^m)^{\a\g} \partial_m A_\g  - 2\,
v_\a W^\a \over v\cdot \lambda}
\right)\,.
}
Thanks to the relation~\ebY\ between the Oda-Tonin $b_Y$-ghost and the
non-minimal $b_\nm$-ghost we deduce that as well $b_{\nm\, 0}(U)= \p^m
A_m+Q_\nm\Omega$.   This  leads to  the  usual  gauge  fixing for  the
Maxwell field which has been derived from the Siegel gauge condition.  As a
consequence of the Siegel  gauge fixing, the Virasoro constraints must
follow and  the vertex  operator is primary.  Indeed, we act  with the
BRST charge from the left on \actB\ and we get
\eqn\actC{
Q_\nm\Big( b_{\nm\,0} (U) \Big) = Q_\nm \Big(\partial^m A_m\Big) \,, 
}
then using the relation $[Q_\nm, b_{\nm\,0}] = L_0$ and $Q_\nm U=0$ we
finally obtain that
\eqn\actD{
L_0 (U) = Q_\nm \Big(\partial^m A_m \Big) 
}
Evaluating the right-hand-side 
\eqn\actE{\eqalign{
Q_\nm \Big(\partial^m A_m \Big) &= \lambda^\a \partial^m \Big( D_\a A_m \Big) = 
\lambda^\a \partial^m  \Big( \p_m A_\a + (\gamma_m)_{  \a\b} W^\b \Big)
\cr
&= 
\lambda^\a \p^2 A_\a + \lambda^\a (\g^m)_{\a\b} \p_m W^\b 
}}
and choosing the gauge $\p^m A_m = 0$, using   $\p^2 A_\a =0$ and the Dirac equation $\not\!\p W =0$, we end up with the Virasoro constraint 
$L_0 (U) = 0$ and the vertex operator is primary. Notice that 
if it were that $L_0 (U) = \rho U$ where $\rho$ is a proportionality constant, then  $U$ would  not be in the cohomology. In addition, it can be proved that, at least on the vertex $U$, $b_{Y 0}$ is nilpotent. 
 
 \subsec{Siegel-Zwiebach gauge for closed strings}

In the case of closed strings, we have a left- and a right-moving $b_\nm$-field that can be used to 
impose the gauge fixing condition. In that case, on the contrary to the
open strings case, the BRST condition 
does  not  impose the  Virasoro  constraints  and  the level  matching
condition.  The  level  matching  condition is  obtained  by  imposing
$b_{0L}  - b_{0R}$  on the  physical  states (where  $L/R$ denote  the
holomorphic  and the  anti-holomorphic part).  See for  example~\refs{\ZwiebachIE}  for a
discussion of these points. 
In the following we will show that imposing the level matching condition leads also to the Virasoro constraints. 

The closed unintegrated vertex operator $U$ is given by the 
expression 
\eqn\cloA{
U = \lambda^\a \lambda^{\hat\a} A_{\a\hat\a}(x, \theta,\hat\theta)\,,
}
where $\lambda^{\hat\a}$ is the pure spinor for the right-moving part. The superfield $A_{\a\hat\a}$ 
depends upon the two supercoordinates $\theta$ and $\hat\theta$. This superfield plays the role of 
the spinorial connection for the supergravity multiplet. In order to relate this superfield to the conventional superfields $A_{mn}$ (whose lowest component is the combination of the metric and 
of the NSNS two form) one needs to derive a ladder of differential equations starting from 
\eqn\cloB{
D_{(\a} A_{\b) \hat\g} = (\g^m)_{\a\b} A_{m \hat\g}\,, 
\quad\quad
\hat D_{(\hat\a} A_{|\b| \hat\g)} = (\g^m)_{\hat\a\hat\g} A_{\a m}\,. 
}
The complete set of equations were derived in~\refs{\GrassiIH}.
Acting with the left- $b_{YL}$ and right $b_{YR}$ Oda-Tonin $b_Y$-fields on the vertex operator~\cloA, we get 
\eqn\cloC{\eqalign{
&(b_{YL, 0} \pm b_{YR,0 }) \Big( 
\lambda^\a \lambda^{\hat\a} A_{\a\hat\a}(x, \theta,\hat\theta)
\Big) \cr
&
= \lambda^{\hat\a} \p^m A_{m \hat\a} + \lambda^\a \p^m A_{\a m} + (Q_{L\nm} + Q_{R\nm})(\Omega)
}}
where $\Omega$ is a polynomial obtained after Fierz rearrangements. As in the open string case, 
these exact terms are irrelevant. Notice that since the right-hand-side involves explicitly the ghost field $\lambda^\a$ and $\lambda^{\hat\a}$, this yields the gauge fixing condition
\eqn\cloD{
\p^m A_{m\hat\a} =0\,, \quad\quad
\p^m A_{\a m} =0\,.}
Using the equations
\eqn\cloE{
\hat D_{(\hat \a} A_{m\hat \b)} = (\g^n)_{\hat\a\hat\b} A_{mn}\,,
\quad\quad
\hat D_{(\a} A_{\a)m} = (\g^n)_{\a\b} A_{nm}\,.
}
By separating  the symmetric and antisymmetric part  of $A_{nm}$ these
equations lead to the usual De Donder gauge for the metric and Landau gauge for the NSNS two form
\eqn\cloF{
\p^m A_{mn} =0\,, \quad\quad
\p^n A_{mn} =0\,.
}
Finally, using
\eqn\cloG{
D_\a A_{mn} - \p_m A_{\a n} = (\gamma_m)_{ \a\b}  W^\b{}_n\,, 
\quad\quad
D_{\hat\a} A_{mn} - \p_m A_{n \widehat\a} = (\gamma_m)_{ \hat\a\hat\b}  W^{\widehat\b}{}_n\,, 
}
where $W^\a{}_n$ is the gravitino superfield.
This implies the set of  equations
\eqn\cloH{
D_\a \p^m A_{mn} - \p^2 A_{\a n} = \not\!\p W_n\,, 
\quad\quad
D_\a \p^n A_{mn} - \p_m \p^n A_{\a n} = (\g_m)_{\a\b} \p^n  W^\b{}_n\,, 
}
Using  Dirac  equation  $\not\!\p  W^\a{}_m=0$ and  the  gauge  fixing
condition $\p^m W^\a{}_m=0$ we obtain that 
$\p^2 A_{\a p} =0$ and  $\p^2 W^\a{}_m =0$. In the same way, one can derive the gauge fixing condition for the other gaugino. The Dirac equation for the gravitino using the present framework
was discussed in~\refs{\GrassiIH}. Notice that unlike the case of bosonic string, we naturally impose both 
conditions  on the vertex  operator $b_{YL,0}$  and $b_{YR,  0}$ since
they depends  upon the independent  left- and right-moving pure  spinor ghosts
that implies the independence of the left- and right-moving $b$-fields. This means that besides the Virasoro constraints also the level matching is automatically imposed. 

\newsec{Regulating the non-minimal pure spinor amplitudes}

Because  the non-minimal $b_\nm$-ghost  has $1/(\l\cdot\lb)$  pole and
measure of  integration over the conjugated ghosts  bring some inverse
powers of  $\l$ and $\lb$ (see  below for details)  the amplitudes can
develop singularities~\refs{\BerkovitsBT,\BerkovitsVI} from the tip of
the pure spinor cone $\l,\lb\sim0$.

In order  to understand the  effect of the  choice of the regulator  on the
amplitudes we analyse  the effect of the general  regulator 
\eqn\ePsiNew{
\Psi =\lb_\a \theta^\a f(\l\cdot\lb)- {1\over2}\,\sum_{I=1}^g S_{mn}^I
{\cal O}^{mn}_I+\sum_{I=1}^g S^I{\cal O}_I 
}
for $f$ is a real function. And ${\cal O}^{mn}_I$ and ${\cal O}_I$ are
ghost number zero $\Lambda$- and $L$-gauge invariant version of~\cNa\ that that will depend on the zero-modes conformal weight
one fields and will be discussed in section~4.4.

\subsec{The vacuum of the pure spinor theory}

The normalisation of the vacuum  of the pure spinor theory $|0\rangle$
is defined by considering its overlap with the  highest ghost number state in
the         zero        momentum         cohomology        $|C\rangle=
(\l\g^m\theta)(\l\g^n\theta)(\l\g^p\theta)(\theta\g_{mnp}\theta)$ 
\eqn\eTree{
\langle 0|C\rangle=\int d^{10}x d^{16}\theta\int [d\l][d\lb][dr]\,\widehat\cN\,
(\l\g^m\theta)(\l\g^n\theta)(\l\g^p\theta)(\theta\g_{mnp}\theta)\, e^{-S_{ps}}\ .
}
with the measures of integrations given in~\edlBerk\ and 
\eqn\edlbr{
[d\lb][dr]=                               d\lb_{\a_1}\wedge\cdots\wedge
d\lb_{\a_{11}}\,\times\,                \partial_{r_{\a_1}}\wedge\cdots
\wedge\partial_{r_{\a_{11}}}\ ,
}
The integration over the pure spinor cone requires that one regulates
the integral. A generic regulator 
\eqn\eNewN{
\widehat\cN=           \exp\left(-           (\l\cdot           \lb)\,
f(\l\cdot\lb)+r_\a M^\a{}_\b\theta^\b\right)
}
where $M^\a{}_\b=\delta^\a{}_\b \, f(\l\cdot\lb)+\l^\a\lb_\b\, f'(\l\cdot\lb)$.
This quantity~\eTree\ gives the normalization of the amplitudes and
the prescription  for evaluating the integration over  the pure spinor
ghosts~\refs{\BerkovitsPX}.

Two detailed evaluations of this integral are given in the appendix~A. 
Setting $h(\l\cdot\lb)=(\l\cdot\lb)f(\l\cdot\lb)$  the amplitude takes
the form
\eqn\eStepFour{\eqalign{
\langle 0|C\rangle&= 11!5!\int \prod_{i=1}^{11} d\l^{\a_i}
d\lb_{\a_i}\,\,(\l\cdot\lb)^{-10}\,e^{-h(\l\cdot\lb)}\,
h(\l\cdot\lb)^{10}\, h'(\l\cdot\lb)
}}
This expression is proportional to 
\eqn\eII{\eqalign{
\langle 0|C\rangle&\propto (-\partial_\alpha)^{10} \int_0^\infty dx \, e^{-\alpha h(x)} h'(x)\Big|_{\alpha=1}\cr
&\propto
 (-\partial_\alpha)^{10}\left(-{1\over\alpha}\,\left(e^{-\alpha\, h(\infty)}-e^{-\alpha\,h(0)}\right)\right)\Big|_{\alpha=1}
}}
We see that the value of the amplitude~\eTree\ is controlled by the value of the
regulator at the boundary of the pure spinor space $\l\cdot\lb=\infty$
and $\l\cdot\lb=0$. Therefore any regulator so that $\lim_{x\to\infty}
 \exp(-h(x))=0$ and $\lim_{x\to0} 
 \exp(-h(x))=0$ is too  strong and will lead to  a vanishing amplitude
 trivializing the theory.\foot{We thank Nathan Berkovits for an important
 discussion concerning this point.}

In the rest of this paper we will make 
the choice of a gauge fermion which is strongly dumped at
zero
\eqn\ePsiMod{
\widehat\Psi= {\lb_\a \theta^\a\over (\l\cdot\lb)^2}
}
and the regulator takes the form

\eqn\eNewN{
\widehat\cN                                                           =
\exp\left[-{1\over(\l\cdot\lb)}-r_\a         \left({\delta^\a{}_\b\over
(\l\cdot\lb)^2} -2 {\l^\a\lb_\b\over (\l\cdot\lb)^3}\right)\,\theta^\b\right]
}
With  this  regulator  any  divergences   from  the  tip  of  the  cone
$\l\cdot\lb\sim0$ will  be regularized by the  exponential factor, and
the  region $\l\cdot\lb\to\infty$  will  be regulated  by the  
powers of $1/(\l\cdot\lb)$ coming from the $r$-zero mode contributions.
For this regulator the amplitude in~\eTree\ is a constant 
\eqn\eZC{
\langle 0|C\rangle=11!\,15\, (4\pi)^{10}\, .
}
 that determines the normalisation of the amplitudes.

\subsec{Tree-level Amplitudes}

The   prescription   for    $N$-point   tree-level   amplitude   given
in~\refs{\BerkovitsBT,\BerkovitsVI} is 

\eqn\eTreeOrig{
{\cal A}^{\tree}_N = \int d^{10}x d^{16}\theta\int [d\l][d\lb][dr]\,   |
 \widehat{\cN}|^2\,
U(z_1)U(z_2)U(z_3)\, \prod_{j=4}^N V(z_i)\, e^{-S_{p.s.}}
}
with the  measures of integrations given  in~\edlBerk\ and~\edlbr. 
\medskip

The advantage of using the regulator~\eNewN\
is that the amplitudes are less diverging at for $\lb\sim\infty$ than at
  $\lb\sim0$. 
Because it is possible to generate $1/(\l\cdot\lb)$-poles of any order
by  inserting  enough  $b_\nm$-ghost  (which happens  at  higher  loop
order~\refs{\BerkovitsBT,\BerkovitsVC}),    but   by    ghost   charge
conservation because  the $b_\nm$-ghost has ghost charge  $-1$ and the
physical vertex  operators appear  at ghost charge  $+1$ or  zero, the
integrand of the amplitude divergence at most like $(\l\cdot\lb)^{11}$
for $\l\cdot\lb\to\infty$.

\medskip

We show that with the regulator~\eNewN\ the amplitudes will converge at
the  boundary  $\l\cdot\lb\sim\infty$  of  the pure  spinor  cone. 

By computing the tree-level amplitude with 3
unintegrated vertex operators and 
$N-3$ integrated  vertex operators  as in~\eTreeOrig, the  11 $r$-zero
modes must  come from the regulator.  Therefore  the integrand becomes
 $d\l d\lb/(\l\cdot\lb)^2$ which converges for
$\l\cdot\lb\to\infty$. Using the  representation with all unintegrated
vertex operators and $N-3$  $b_\nm$-ghost insertions, from seven point
$N\geq7$  it is possible  to saturate  the 11  $r$-zero mode  from the
$b_\nm$-ghost only and the integrand
seems to behaves as $d\l d\lb/(\l\cdot\lb)$ which
corresponds to a logarithmic singularity at infinity. But as remarked
in~\refs{\BerkovitsVI} all the terms in the $b_\nm$-ghost commute with
the conserved charges 
\eqn\eQQ{
q_1=\oint \,(r_\a s^\a - \l^\a w_\a);\qquad q_2=\oint \,\lb_\a s^\a\, ,
}
 which imply that the terms of the $b_\nm$-ghost~\nnA\
have opposite $r$-charge and $\l$-charge and are
invariant  under  the  shift   symmetry  $\delta  r_\a=c\,
\lb_\a$ where $c$ is a constant.  Therefore to saturate all the 11 $r$-zero modes we need to
pick 12  $r$ from the $b_\nm$-ghost  or 11 $r$  from the $b_\nm$-ghost
and one $r$ from the regulator and contract the left over $r$-ghost with $s$-ghosts.  In either case this brings enough
powers of $1/(\l\cdot\lb)$
so that the integral converges for the large values of the pure spinor ghost.

\bigskip

In the non-minimal formalism it is possible to construct the following
quantity   $\xi=(\lb\cdot\theta)/(\l\cdot\lb+r\cdot\theta)$   so  that
$Q_\nm\xi=1$.  If this state is allowed it will trivializes the theory
by making all physical state Q-exact, and all amplitude vanishing.  
By  evaluating the amplitude  $\langle 0|\xi\rangle$  we see  that the
contribution with 11 $r$-zero mode lead to a logarithmic divergence at
infinity.  Because the  terms  in the  expansion  of the  $\xi$ do  not
commute with conserved  charges $q_1$ and $q_2$ and  the divergence is
not protected by the symmetry $\delta r_\a=c \,
\lb_\a$. Therefore the state $\xi$ is not allowed in the physical
Hilbert space of the theory.

\medskip

We     can    compare     with    the     prescription     given    by
Berkovits in~\refs{\BerkovitsBT} where the following gauge fermion and
regulator are used

\eqn\nnC{
\Psi= \lb_\a\theta^\a; \qquad \cN= \exp\Big(-\l\cdot\lb - r\theta\Big)\ .
}

The regulator~\eNewN\ takes the form given in~\eNewN\ with 
\eqn\eM{
M^\a{}_\b = { \delta^\a{}_\b\over (\l\cdot\lb)^2} -2 {\l^\a
\lb_\b\over (\l\cdot\lb)^3}
}
This  matrix  satisfies  the  property  that  $M^\a{}_\b  \,  M^\b{}_\g
=\delta^\a{}_\g/(\l\cdot\lb)^4$ that implies that $(M^{-1})^\a{}_\b=
(\l\cdot\lb)^4 \, M^\a{}_\b $.

In the amplitude one can eliminate the dependence on this matrix in the regulator by
performing    the    change     of    variable    $r_\a    =\widetilde  r_\b
(M^{-1})^\b{}_\a$. This induces a non-trivial Jacobian factor depending
only on the $\l$ and $\lb$ pure spinor ghosts 
\eqn\eJJ{
[d\lb][dr]\to          d\lb_{\a_1}\wedge  \cdots\wedge       d\lb_{\a_{11}}\wedge
\p_{\widetilde r_{\b_1}}\wedge \cdots\wedge \p_{\widetilde r_{\b_{11}}}\, \prod_{i=1}^{11}
M^{\a_i}{}_{\b_{i}} \ .
}
We  should stress  here that  this transformation  preserves  the pure
spinor  conditions   since  $\widetilde{\lb}\g^m\widetilde{\lb}=0$  and
$\widetilde{\lb}\g^m \widetilde r=0$.
Because $M^\a{}_\b= \p \widetilde{\lb}_\b/\p_{\lb_\a}$ this Jacobian factor is exactly the one for the transformation
\eqn\etlb{
\widetilde{\lb}_\a = {\lb_\a\over (\l\cdot\lb)^2}\ ,
}
therefore the measure of integration over the pure spinor ghost with the
regulator~\eNewN\ takes the form
\eqn\eNNN{
\int   [d\l][d\lb][dr]  \,   e^{-{1\over(\l\cdot\lb)}-rM\theta}=  \int
[d\l][d\widetilde{\lb}][d\widetilde r] \, e^{-\l\cdot\widetilde{\lb} -\widetilde r\theta}
}
which  is   the  original  regulator~\nnC\   introduced  by  Berkovits
in~\refs{\BerkovitsBT} expressed in terms of the inverted variables.
This shows that  our regulator is making the  pure spinor $\l$ massive
using $\widetilde{\lb}$ instead of $\lb$.

 The massless vertex operators  do not depend on the non-minimal
variables.  This shows  that the tree-level  amplitudes defined with
only three unintegrated vertex operators~\eTreeOrig\ are the same with the 
regulator~\eNewN\ and
the regulator introduced in~\nnC\ in~\refs{\BerkovitsBT}.  

Remarking that

\eqn\eTT{\eqalign{
\lb\g^{mnp}M^{-1}\widetilde      r      &=      (\l\cdot\lb)^2      \,
(\lb\g^{mnp}\widetilde r)\cr
 \widetilde  r_{\b_1} \widetilde  r_{\b_2} \,(M^{-1})^{\b_1}{}_{[\a_1}
 \,(M^{-1})^{\b_2}{}_{\a_2}   \,  \lb_{\a_3]}  &=   (\l\cdot\lb)^4  \,
 \widetilde r_{[\a_1} \widetilde r_{\a_2} \, \lb_{\a_3]}
}}
and introducing $s^\a=\widetilde s^\b\,
M^\a{}_\b$ the $b_\nm$-ghost transforms as 
the non-minimal $b_\nm$-ghost of eq.~\eBB\ transforms as
\eqn\ebTr{\eqalign{
b_\nm &=\widetilde s\p \widetilde{\lb} + {1\over 4}\, {{\widetilde{\lb}}_\a\over \l\cdot{\widetilde{\lb}}}\, {\bf b}^\a\cr
{\bf    b}^\a    &\equiv    G^\a+     {{\widetilde    r}_\b    \over
(\l\cdot{\widetilde{\lb}})} H^{\a\b}+
{\widetilde r_\b \widetilde r_\gamma\over(\l\cdot{\widetilde{\lb}})^2}\,
K^{\a\b\gamma}+ {\widetilde r_\b \widetilde r_\g \widetilde r_\delta \over(\l\cdot{\widetilde{\lb}})^3}\, L^{\a\b\gamma\delta}
}}
Since   the    operators   $G^\a$,   $H^{\a\b}$,    $K^{\a\b\g}$   and
$L^{\a\b\g\delta}$ do not depend on the non-minimal sector, this shows
that this expression is identical to the one in~\eBB\ and shows the equivalence of the amplitudes with the insertion of
the $b_\nm$-ghost.   It is important that the  $b_\nm$-ghost keeps the
same functional dependence in the $\widetilde r_\a$, $\widetilde s^\a$
and $\widetilde{\lb}_\a$
variables are in the $r_\a$, $s^\a$ and $\lb_\a$ variables.

\medskip

Because we are not transforming  the conjugated ghost $w_\a$ and $\bar
w^\a$ and  because the $\Lambda$-  and $L$-gauge invariant  measure of
integration  over these  variables bring  inverse powers  of  the pure
spinor  ghost, we will  show that  this regulator  provides divergence
free amplitudes that converge at $\l,\lb\sim \infty$.

\subsec{Regulating the higher-loop amplitudes}

The prescription for a genus-$g$  amplitude in this formalism is given
by \refs{\BerkovitsBT}

\eqn\ePmn{
{\cal A}^g_N =\int d^{3g-3}\tau\,\left\langle \left| \prod_{i=1}^{3(g-1)}(\mu_i|b_{\nm})\,
\widehat\cN\, \right|^2\prod_{i=1}^N V_i\right\rangle\ .
}
The       integration      over       the       world-sheet      field
$[x^m,p_\a,\theta^\a,\l^\a,w_\a,\lb_\a,\bar     w^\a,r^\a,s_\a]$    is
defined by 
\eqn\eIntNN{
\left\langle \cdots \right\rangle = \int d^{10}x d^{16}\theta \int
[d\l][d\lb][dr]\prod_{I=1}^g\int [dw^I][d\bar w^I][ds^I]\,\cdots e^{-S_{ps}}
}
The measure of integration over the conjugated ghosts is given by 
\eqn\edwwww{\eqalign{
[dw^I]&=M^{\a_1\cdots \a_8}_{m_1n_1\cdots
m_{10}n_{10}} dN^{m_1n_1\, I}\cdots dN^{m_{10}n_{10}\,I} dJ^I\,\p_{\l^{\a_1}}\cdots\p_{\l^{\a_8}}\cr
[d\bar w^I][ds^I]&=\prod_{i=1}^{10}  d\bar  N_{m_in_i}^I\wedge
d\bar J^I \wedge \prod_{i=1}^{10} \partial_{S^I_{m_in_i}}\wedge \partial_{S^I}
}}
where we set   $M_{m_1n_1\cdots   m_{10}n_{10}}^{\alpha_1\cdots   \alpha_8}={ 
(\g_{m_1n_1m_2m_3m_4})^{((\a_1\a_2}
(\g_{m_5n_5n_2m_6m_7})^{\a_3\a_4}(\g_{m_8n_8n_3n_6m_9})^{\a_5\a_6} }$ 
${(\g_{m_{10}n_{10}n_4n_7n_9})^{\a_7\a_8))}}$
and   $((\cdots))$   means   that   one  considers   the   symmetrized
$\gamma$-traceless part.

In  order  to regulate  the  integration over  the  zero  mode of  the
conjugated ghosts we make the following choice
\eqn\eOO{\eqalign{
{\cal O}^{mn}_I  &= (w_I\g^{mn}\l) \cr
{\cal O}_I &= (w_I\l)\ .
}}
The zero modes are defined  by   integration  over  the  homology  $a$-cycles
$\Phi^I\equiv \oint_{a_I} \Phi$ for $1\leq I\leq g$.

 The associated regulator 

\eqn\newnnCBA{\eqalign{
\cN &=\exp{[Q_\nm,\Psi]}\cr
&    =    \exp\Big[   -    {1\over\l    \cdot\lb}   -    {r\cdot\theta\over
(\l\cdot\lb)^2}+2{(\lb\cdot\theta)(r\cdot\l)\over (\l\cdot\lb)^2}\Big]\cr
&\times\exp\Big[ -
 \sum_{I=1}^{g} \Big({1\over2} N^I_{mn} \bar N^{I\,mn}+J_I\bar J^I\Big)\Big]
 \cr
&\times\exp\Big[-\sum_{I=1}^g\, {1\over4} S^I_{mn} (d^I\g^{mn}\l) 
+ S^I (\l d^I) 
\Big]\cr
}}
The third  and fourth line  are the $\Lambda$ and  $L$-gauge invariant
version of the regulator $\exp(- w\cdot\bar w- s\cdot d)$.

\subsec{Zero mode counting in the non-minimal formalism}

For having a non vanishing massless $n$-point genus $g$ amplitude one needs to satisfy the
fermionic zero modes constrains given by the following equations
\eqn\esysZM{\eqalign{
11g &= n_{ds}+n_{s\partial\lb}\cr
11&= n_{r\theta}+n_{rd^2}+n_{rd^0}+2n_{r^2d}+3n_{r^3d^0}\cr
16g&= n_{ds}+n_{d\,vop}+n_{r^0d}+2n_{rd^2}+n_{r^2d}\cr
3(g-1)&=n_{s\p\lb}+n_{r^0d}+n_{rd^2}+n_{rd^0}+n_{r^2d}+n_{r^3d^0}\cr
}}
where  $n_{ds}$  is  the  number  of $S\l  d$  contributions  from  the
regulator, $n_{r\theta}$ is the number of $r\cdot\theta$ contributions
from the  regulator, $n_{d\, vop}$  is the number of  $d$ contributions
from the vertex operators and $n_{s\p\lb}$ and $n_{r^id^j}$ with $(i,j)\in\{(0,1),(1,0), (1,2),
(2,1), (3,0)\}$ are the various contribution from the $b_\nm$-ghost.

The $d$-zero mode constraint implies that
\eqn\edZM{
2g = n_{d\,vop}+n_{rd^2}-n_{r^3d^0}-3-2n_{s\p\lb}
}
Since $n_{rd^2}\leq 11$ and  $n_{r^3d^0}\geq0$ we deduce that this system
of equation does not have a solution after genus
\eqn\eGenusMax{
g> {1\over 2}\, n_{d\,vop}+4\ .
}
An $n$-point massless amplitude would vanish for all genus $g\geq 5+n/2$ if
there are no singularities in the pure spinor integration.

With the $1/(\l\cdot\lb)$ regulator introduced in the previous section
the integration over the pure spinor ghost $\l$ and $\lb$ behaves as

\eqn\ellInt{
I_{\l,\lb}=\int_0^\infty       {d\l      d\lb\over       \l\cdot\lb}       \,      {1\over
(\l\cdot\lb)^{n_{r\theta}}}\,
\left(\l\cdot\p\lb\over\l\cdot\lb\right)^{n_{s\p\lb}}\, e^{-{1\over\l\cdot\lb}}
}
the  $q_1$ and  $q_2$  invariance of  the  $b_\nm$-ghost implies  that
$n_{r\theta}+n_{s\p\lb}\geq1$  and   these  integrals  are   converging  both  at
$\l\cdot\lb=0$ and $\l\cdot\lb\sim\infty$.

\medskip

This analysis  shows that the $b_\nm$-ghost and the vertex
operators do not provide  enough fermionic zero mode contributions for
having  non vanishing  amplitudes at  high enough  genus  order, which
is incompatible with unitarity. 

Therefore unless there are extra sources of $d$-zero modes the theory cannot be
unitary.

\medskip

Before presenting  a possible solution to this  problem in section~4.5
we make a few comments on the heat kernel regularisation.

\bigskip
\noindent{\sl The heat kernel regularisation~\refs{\BerkovitsVI,\BerkovitsNEW}}

A  heat kernel  regularisation of  the pure  spinor  singularities was
introduced in~\refs{\BerkovitsVI}.  When  the amplitude develops
higher-order divergences  with $11n< n_{r\theta}<  11(n+1)$ one should
add~\refs{\BerkovitsNEW}   $n$  sets   of   regulating  pure   spinors
$(f_\a,\bar f^\a,g^\a, \bar g_\a)$ where $f^\a$ and $\bar f_\a$ are bosonic constant pure spinors and
$g^\a$  and $\bar  g_\a$  are fermionic  constant  pure spinors. 
Each set of regulators is integrated over according the prescription (see
equations~(3.20) and~(3.29) of~\refs{\BerkovitsNEW})
\eqn\eNewHeat{
\int d^{11}f d^{11}\bar f d^{11}g d^{11}\bar g\, e^{\sum_{I=1}^g \, (f^\a w_{\a\, I} + g^\a
d_{\a\, I}+\bar f_\a \bar w^\a_I+\bar g_\a s^\a_I)}\ ,
}
Each  extension can provide $n_{\bar g s}$ extra $s$-zero mode and
$n_{gd}$ extra $d$-zero modes contributions to the counting in~\esysZM

\eqn\esysZMII{\eqalign{
11g &= n_{ds}+n_{s\partial\lb}+n_{\bar g s}\cr
16g&= n_{ds}+n_{d\,vop}+n_{r^0d}+2n_{rd^2}+n_{r^2d}+n_{gd}\cr
}}
leading to the $d$-zero mode counting 

\eqn\edZMII{
2g = n_{d\,vop}+n_{rd^2}-n_{r^3d^0}-3-2n_{s\p\lb}+n_{gd}-n_{\bar g s}
}
Since $n_{\bar g s}\geq 0$ and $n_{g\,d}\leq11\, n$, where $n$ defined
by the
order of the $\l\cdot\lb$ pole, we deduce that this system
of equation does not have a solution after genus
\eqn\eGenusMax{
g> {1\over 2}\, n_{d\,vop}+4+{11n\over2}\ .
}
In particular with one set $n=1$ of regulator the massless four-point amplitudes
  will  be  vanishing after  genus
$g\geq  12$, which  would  not  be compatible  with  unitarity if  the
amplitudes did not had any divergences.     In order
that the massless four-point amplitude does not vanish after some loop order
one needs to have a degree of divergence that increases with the genus
order. Poles are generated by picking extra $r$-field
 from the $b_\nm$-ghost.  Because the number of $r$-zero
mode is  at most 11, these higher  order pole can only  arise from the
non-zero mode part  of the $r$-field and their  contraction with extra
$s$-fields provided by the regulator factor in~\eNewHeat.  
 The same issue arises by increasing the number of external legs at a given genus orderwhen using a representation of the
 amplitudes with  unintegrated vertex  operators.

 In order
that the  $d$-zero mode saturation can  be satisfied to  all orders in
perturbation one needs that $n_{g\,d}-n_{\bar g\,s}>0$ and that this quantity
increases (may be not
monotonically) with the genus order and the number of punctures.
As well, with the necessity of introducing many regulating set, one could be
worry that the multi-dimensional integration over the $f_\a$ and $\bar
f^\a$  pure  spinors variables  leads  to  extra  poles at  unphysical
positions. 
In order to avoid adding more and more regulators one can consider\foot{  We  would  like  thank  Nathan
Berkovits and  Yuri Aisaka for  this suggestion.}  introducing an
infinite set   from  which  only  a  finite subset  will  contribute  to  the
amplitudes at a given order. It would be interesting to clarify these points.

\medskip

  It could be  interesting to relate this approach to  the one used in this
present work, and it is tempting to conjecture that the infinite set of
regulators  setup can   be related  to the  field redefinition
introduced in~\etlb.

\subsec{Adding $d$-zero mode contributions}

In  order to  resolve the  issue of  the vanishing  of  the amplitudes
because  of the  impossibility of  saturating all  the  $d$-zero modes
after some genus order, we introduce the following piece to the gauge fermion
\eqn\ePsidd{
\widehat\Psi=\Psi+\sqrt{\alpha'}\, \sum_{1\leq I,J\leq g} S^I_{mn}\, (d^I \g^{mnp} d^J)\, P_{p\, J}
}
which modifies the regulator as
\eqn\newnnCB{\eqalign{
\widehat{\cN} &= \cN\times \cN_{d}\cr
\cN_d&=\exp\Big[ -\sqrt{\alpha'} \, \sum_{1\leq
I,J\leq g} \bar N^{mn}_{I} \, (d^{I}\g_{mnp}d^{J}) \, P^p_J\Big]\cr
&\times\exp\Big[-\sqrt{\alpha'}\sum_{1\leq
I,J\leq g} S^{mn}_{I} \, \left(P^I_s\, (\l\g^s\g_{mnp}d^{J}) \,
P^p_J+ (d^{I}\g_{mnp}\g^s\l) \,P^J_s
P^p_J\right)\Big]\cr
}}
With this addition to the  regulator the $d$-zero mode counting in the
$n$-point  amplitude at genus  order $g>4+n/2$  can be  satisfied by
picking $g-(4+n/2)$ contributions of $N^{mn}_I \, (d^{I}\g_{mnp}d^{J})
\, P^p_J$.

Under  the change of  variables $\lb\to\widetilde{\lb}$  of eq.~\etlb,
the extension of the gauge fermion in~\ePsidd\ transforms as 
\eqn\ePsiddII{
\delta\Psi = -\sqrt{\alpha'}\, \sum_{1\leq I,J\leq g} {\widetilde S}^I_{mn}\, (d^I \g^{mnp} d^J)\, P_{p\, J}
}
where           ${\widetilde           S}^I_{mn}=           {\widetilde
r}^I\g_{mn}\widetilde{\lb}$.   But   only  the  second   line  of  the
regulator $\cN_d$  is invariant. This  implies that this  extension of
the  regulator makes  a difference  between the  non-minimal formalism
regulated with  a mass $\l\cdot\lb$  introduced in~\refs{\BerkovitsBT}
or the mass $\l\cdot\widetilde{\lb}$ used here.

We   could  not   justify  this   extension  by   a   first  principle
derivation. The difficulty of  saturating $d$-zero mode at higher loop
could be related to a  background charge screening constraint which is
not immediately visible, except for  the vanishing of certain class of
amplitudes,  due to  the gauge  fixed  definition of  the pure  spinor
formalism.

\newsec{Multigraviton amplitudes at higher-loop }

 The closed string massless vertex operators is defined as \refs{\BerkovitsBT,\BerkovitsVI}

\eqn\eVop{
V=\int d^2z \,\left(G_{MN}(X) \partial X^M \bar\partial X^N+W^{\a\b}\,d_\a\widehat d_\b+\cdots \right)
}
where  $X^M=(x^m,\theta^\a,\widehat\theta^\ha)$, the  symmetric part
of $G_{(MN)}$  is the graviton  superfield and the  antisymmetric part
$G_{[MN]}$           is           the           NS           $B$-field
superfield.    $W^{\a\b}(x,\theta^\a,\widehat\theta^\ha)$   is   the
dimension one gauge-invariant superfield whose lowest component is the
Ramond-Ramond field strength.

The zero modes saturation of a $n_{grav}$-graviton amplitude at genus  $g\geq2$  leads to
\eqn\nnD{\eqalign{
{\cal A}_N^g &= \int d^{3g-3}\tau\,\left\langle \Big| \prod_{i=1}^{3g -3} 
( \mu_i |b_{\nm} ) \,  \widehat\cN  \Big|^2 \, V^{n_{grav}}\right\rangle \cr
\sim& \Big\langle   \Big|
 e^{-{1\over\l\cdot\lb}-\sum_{I=1}^g N_I\cdot \bar N^I}\cr
&\times
\Big({\chr \theta\over\l\cdot\lb}\Big)^{n_{r\theta}} 
\Big(S \l d\Big)^{n_{sd}}  \Big(\bar N d^2 P\Big)^{n_{d^2P}}\,
\Big(\p\theta A + \Pi B + d W + N F\Big)^{n_{grav}}  \cr
\times&
\Big(s \p \bar\l\Big)^{n_s} 
\Big(\chl\Pi d\Big)^{n_{r^0d}}
\Big(\chl \, \chr\, d^2 \Big)^{n_{rd^2}} 
\Big(\chl \, \chr\, N\Pi\Big)^{n_{rd^0}}
\Big( \chl \chr^2 d N\Big)^{n_{r^2d}} 
\Big( \chl \chr^3 N^2 \Big)^{n_{r^3d^0}}\Big|^2\Big\rangle
}}
where we made use of the variables~\eChLr.
We have schematically written down all possible terms coming from the regulator $\widehat\cN$ and the
$b_{\nm}$-ghost using the notations of eq.~\eChLr.  When $n_{r\theta}$ is non zero the contribution
is given by an integrations over a subspace of the $\theta$-superspace
 but when $n_{r\theta}=0$ this is a full superspace integral. 
The various powers in~\nnD\ satisfy the   constraint
\eqn\bA{
3g -3 = n_s+ \sum_{i=0}^{4} n_i\,,
}
 that   there are $3g-3$  insertions of the (left-moving)  $b_{\nm}$-ghost.
The saturation of the  $11g$ $s^\a$-zero modes, the $16g$ $d_\a$-zero modes, and the $11$ $r_\a$-zero mode gives
\eqn\esCount{\eqalign{
&s:~~~~11g = n_s+ n_{sd} -n_{s,r}\cr
&d:~~~~16g = n_{sd} +2n_{d^2P}+ n_{r^0d} + 2 n_{rd^2} + n_{r^2d}+n_{grav}\cr
&r:~~~~11 = n_{r\theta}  + n_{rd^2} + n_{rd^0} + 2 n_{r^2d} + 3 n_{r^3d^0}-n_{s,r}\,.
}}
where $n_{s,r}$  is the number  of contractions between  the $s$-ghost
and the $r$-ghost. 

\subsec{The four-graviton amplitude at higher-genus $g\leq 6$} 

For the  case of the  four-graviton amplitude, with  $n_{grav}=4$, the
previous  conditions have  the  following solution  valid until  genus
$g\leq 6$ \refs{\BerkovitsVC}
\eqn\eSolveB{\eqalign{
&n_{r\theta}=12-2g,\quad n_{sd}=11g,\quad n_{rd^2}=2g-1,\quad n_{r^0d}=g-2 \cr
&n_s=n_{rd^0}=n_{r^2d}=n_{r^3d^0}=n_{d^2P}=n_{s,r}=0\ ,
}}
 which   corresponds   to  the   partial   superspace  integral   when
 $n_{r\theta}=12-2g\neq0$  giving the leading  contribution to  the low-energy
 limit of the string amplitude \refs{\BerkovitsVI,\BerkovitsVC} 
\eqn\eGenusg{
{\cal      A}_4^g\sim      \int     d^{16}\theta      d^{16}\bar\theta
\theta^{12-2g}\bar\theta^{12-2g}
(W_{\alpha\beta})^4\times I^g\sim (\alpha'\partial^2)^g R^4\, I^g+O(\alpha' k^2)\ .
}
where  $I^g$ is  a field  theory integral  which is  the low-energy
energy of the expression arising from the integration over the moduli.

For  this case   the good convergence properties
over the spinor variables allowed to perform the change of variables
$\l\to\widetilde{\lb}$ of  eq.~\etlb\ and  use the BRST  invariance to
set $\cN_d=1$.  By using the same  steps as in section~4.2  we can map
our amplitude computation to the one in~\refs{\BerkovitsVC}
leading to  identical results.

For the solution~\eSolveB\ the form of the integrand is given by 

\eqn\eIg{
I^g=\int d^{3g-3}\tau\, \prod_{i=1}^4\int d^2z_i
e^{ik^i\cdot x}(z_i)\, \left|\prod_{i=1}^{3g-3}\int d^2y \mu(y_i)\,
\prod_{j=1}^{g-2}\Pi(y_j)\right|^2
}
The expression involves $2(g-2)$ insertions of the supersymmetric loop
momenta $\Pi^m\sim  \partial x^m+ (\theta\g^m\partial\theta)/2$  flowing through the
loops.  The field theory limit of this amplitude in ten dimensions has
$3g-3+4=3g+1$ propagators,  and $2(g-2)$ are  loop momentum contracted
between themselves or to external polarisation or some of the explicit
external  momenta   in~\eGenusg.  The  resulting   integral  has  mass
dimension $(D-4)g-6$ as it should be by dimensional analysis. Such an
expression displays the explicit superficial ultra-violet behaviour of
the amplitude.

\subsec{The four-graviton amplitudes at higher-genus $g\geq7$}

At  genus  $g\geq7$  the  massless  four-point  amplitude  can  develop
divergences  in the pure  spinor integration at the  tip of  the cone
$\l\cdot\lb\sim0$~\refs{\BerkovitsVC}, and the change of variables
$\l\to\widetilde{\lb}$ of eq.~\etlb\ is not allowed. 
 As well because of
the potential divergences in the pure spinor integration we cannot use
the BRST invariance to set $\cN_{d}=1$. We will see that this extra
contribution to the $\cN_{d}$ regulator will bring extra $d$-zero
mode allowing the saturate the fermionic zero mode after $g\geq7$.   Because the new contributions  to the regulator
 come with one power of  $\alpha'$ we want  to minimize the
number of terms  coming from this modification of  the regulator to get
the leading contribution to the low-energy limit of the amplitude.
This is accomplished by the solution parametrized 
\eqn\eFSol{\eqalign{
n_s&=1,\quad n_{r^0d}=3g-14, \quad n_{rd^2}=12, \quad n_{rd^0}=n_{r^2d}=n_{r^3d^0}=0,\cr
n_{r\theta}&=0,\quad n_{sd}=11g,\quad  n_{d^2P}=g-6\ .
}}
where we have taken $n_{rd^2}>11$ $r$-zero mode from the $b_\nm$-ghost
as required by the invariance under the charges~\eQQ.

This expression leads to a low-energy expansion of the four-graviton amplitude
in ten dimensions

\eqn\eGenusgNew{
{\cal      A}_4^g\sim  (\alpha')^{g-6}\,    \int     d^{16}\theta      d^{16}\bar\theta
(W_{\alpha\beta})^4\times \tilde I^g\sim (\alpha')^g\partial^{12}R^4\, \tilde I^g+O(\alpha' k^2)\ .
}
where now $I^g$ is 
\eqn\eIgNew{\eqalign{
\tilde I^g=\int d^{3g-3}\tau\, \prod_{i=1}^4\int d^2z_i
e^{ik^i\cdot x}(z_i)\, \left|\prod_{i=1}^{3g-3}\int d^2y \mu(y_i)\,
\prod_{j=1}^{g-2}\Pi(y_j)\, \prod_{I=1}^{g-6}\oint_{a_I}\Pi\right|^2
}}
because  this expression  contains $2g-8$  powers  the supersymmetric
loop momenta running in the loop, this expression has mass dimension $(D-2)g-18$
and  taking into  account  the dimension  twenty operator  $\p^{12}R^4$
multiplying  the  amplitude the  total  amplitude  has mass  dimension
$(D-2)g+2$.  This confirms this is the
leading  contribution   to  low-energy  limit   of  the  four-graviton
amplitude in ten dimensions. 

In the extreme case that all the $g-6$ powers of loop momenta from the regulators are
contracted with  plane-wave factors, the amplitude with  have an extra
factor of $2(g-6)$  powers of external momenta and  will behaves as

\eqn\eAmpMost{
{\cal A}_4^g\sim \int d^{3g-3}\tau\, \left|\prod_{i=1}^{3g-3} d^2y \mu(y_i)\,
\prod_{j=1}^{g-2}\Pi(y_j)\right|^2\, {\alpha'}^g\partial^{2g}R^4+O((\alpha'k^2)^{g+1}R^4)
}
For  this contribution  to  be  the leading  low-energy  limit of  the
$g$-loop   four-graviton   amplitude  at   genus   order  $g>6$   many
cancellations within the  integrals~\eIgNew\ beyond the supersymmetric
ones  must take place.   They could  be the  consequence of  the extra
cancellations detailed in~\refs{\BjerrumBohrJI,\BadgerRN} occurring in
the on-shell colorless amplitudes.

\subsec{Vanishing of $N<4$-point amplitudes}

Since  the regulator~\newnnCB\  or the  regularized $b_\epsilon$-ghost
of~\refs{\BerkovitsVI} bring an arbitrary number of $d$-zero modes one
needs to make  sure that all massless $N$-point  amplitudes with $N<4$
vanish to all order in perturbation.  The vanishing of the $N<2$-point
amplitudes  imply  by  factorisation  and the  absence  of  unphysical
singularities   in   the   amplitude,   the   finiteness   of   string
perturbation~\refs{\Atick,\Martinec,\lechtenfeld,\BerkovitsHG}.     The
vanishing of  the 3-point amplitude  at higher genus is  not necessary
for the finiteness  of string perturbation but is  a necessary but not
sufficient condition  for the absence of  infra-red singularities when
taking the low-energy limit of the four-point string amplitudes in ten
dimensions.

It was  shown in~\refs{\BerkovitsPX} that  in the minimal  pure spinor
formalism  all  the $N<4$-point  amplitudes  vanish  to  all order  in
perturbation.

The vanishing of the vacuum diagram is ensured by the integration over
the sixteen left-moving and right-moving superspace variables. For the
following argument  we will assume  that all the vertex  operators are
unintegrated. The vanishing of  the 1-point amplitude is a consequence
of the on-shell relation. At most the integrand can bring 11 powers of
$\theta$  and  the  amplitude   takes  the  form  $\int  |d^{16}\theta
\theta^{11}  |^2\, V_1$  where  $V_1=|(\l\g^m\theta) a_m(x)+(\l\g^m\theta)
(\theta\g_m\chi)+\cdots|^2$ is  a massless vertex  operators where the
ellipsis  are  for  higher-derivative  contributions.   But  one-point
on-shell amplitudes  have $k_1=0$ and  all higher order term  in $V_1$
drops  out  and  the  integral  vanishes after  integration  over  the
$\theta$ variables.  The vanishing  of the two-point amplitude follows
the   same  argument   that  the   integration  over   the  superspace
$\theta$-variables leads to contributions that vanish on-shell because
there is only one on-shell independent momentum.

For the case of the massless three-point function we find that using the
original regulator~\newnnCB\ that the
zero mode constraint can be satisfied for all genus from $g\geq3$. 
But  we will show  that because  all the  contribution have  more than
two-derivative (there  is no renormalisation  of the Planck  mass) the
on-shell condition assure the vanishing of these amplitudes.
 For
the    massless    three-point    amplitude   momentum    conservation
$k_1+k_2+k_3=0$ and the
on-shell conditions $k_1^2=k_2^2=k_3^2=0$  imply that $k_i\cdot k_j=0$
for all $i,j=1,2,3$.
At
genus 3 we have the contribution $n_{rd^2}=6$, $n_{r\theta}=5$ $n_{sd}=33$ and all the
other  integers being  zero  and  three $d_\a  W^\a$  from the  vertex
operators.  
In the  case one picks the 11 $r$-zero mode from the regulator one gets
\eqn\eThreeTwo{
\int  |d^{16}\theta  \theta^{11}|^2  V_1  V_2  V_3\sim  k^2  \,
\hat R^3 +\cdots
}
which means that  one must distribute two momenta  on three powers
of  linearized Riemann  tensor $\hat  R_{mnpq}= k_{[m}\,  \zeta_{n][p} \,
k_{q]}$. This vanishes by the on-shell conditions.
In the case where there is no contributions of $r$-zero mode  from the
regulator one get and amplitudes of the type
\eqn\eThreeOne{
\int |d^{16}\theta |^2 V_1 V_2 V_3\sim k^{13} \hat R^3+\cdots
}
which has more  powers of momenta to contract  and this vanished after
using the on-shell  conditions. The same conclusion is  reached to the
contribution involving the supersymmetric partner of the graviton. This show that the massless 3-point
amplitude vanish to all order in perturbation.

We hope that our considerations help to a better understanding of this
intricate and interesting new 
field. Higher-loop and multileg computations are important for several checks in string 
perturbation theory and beyond, but  in addition, they are needed test
of the soundness of the formalism.

{\bf  Acknowledgements:}  We would  like  to  thank Nathan  Berkovits,
Michael Green, Alberto Lerda, Massimo Porrati and Stefan Theisen for
useful  discussions and  comments  on a  preliminary  version of  this
draft, and Yuri Aisaka and Carlos Mafra for comments on the first version of this preprint.
 We would like to thank the organizers of the KITP
workshop ``Fundamental Aspects of Superstring Theory'' for providing a
stimulating atmosphere where most of this work has been done.
This research was supported in part  (PV)  by the ANR grant BLAN06-3-137168.
This research was supported in part by the National Science Foundation under
Grant No. PHY05-51164.

\appendix{A}{The tree-level amplitude}

We consider the general form of the regulator 
\eqn\ePsiNew{
\Psi =\lb_\a \theta^\a f(\l\cdot\lb)
}
where $f$ is a real function.
With this choice of gauge fermion we have the following regulator
\eqn\eNewnN{
\widehat\cN=           \exp\left(-           (\l\cdot           \lb)\,
f(\l\cdot\lb)+r_\a M^\a{}_\b\theta^\b\right)
}
where $M^\a{}_\b=\delta^\a{}_\b \, f(\l\cdot\lb)+\l^\a\lb_\b\, f'(\l\cdot\lb)$.
With this regulator we evaluate the tree-level integral

\eqn\eTree{
\langle 0|C\rangle= \int d^{16}\theta \int [d\l][d\lb][dr] \,\widehat\cN\, (\l\g^m\theta)(\l\g^n\theta)(\l\g^p\theta)(\theta\g_{mnp}\theta)
}
By performing the integration over the 11 $r$ variables and using that
$(\lb\cdot\theta)^2=0$ we get
\eqn\eStepOne{\eqalign{
\langle 0|C\rangle&= \int  d^{16}\theta \int [d\l]d\lb_{\a_1}\wedge \cdots
\wedge
d\lb_{\a_{11}}\,e^{-(\l\cdot\lb)\,f(\l\cdot\lb)}\,  (\l\g^m\theta)(\l\g^n\theta)(\l\g^p\theta)(\theta\g_{mnp}\theta)\times\cr
&\times 
f(\l\cdot\lb)^{10}                                                   \,
\theta^{\a_1}\cdots\theta^{\a_{10}}\theta^{\sigma}\left(\delta_\sigma^{\a_{11}}\,
f(\l\cdot\lb) +11\, \lb_\sigma \, \l^{\a_{11}} f'(\l\cdot\lb)\right)\ .
}}
Performing the integration over the sixteen $\theta$ variables leads to
\eqn\eStepTwo{\eqalign{
\langle 0|C\rangle&=  \int [d\l]d\lb_{\a_1}\wedge \cdots
\wedge
d\lb_{\a_{11}}\,e^{-(\l\cdot\lb)\,f(\l\cdot\lb)}\,
(\g^m\l)_{r_1}(\g^n\l)_{r_2}(\g^p\l)_{r_3} (\g_{mnp})_{r_4r_5}\times\cr
&\times 
f(\l\cdot\lb)^{10} \, \epsilon_{16}^{\a_1\cdots\a_{10}\sigma r_1\cdots
r_5}\left(\delta_\sigma^{\a_{11}}\,
f(\l\cdot\lb) +11\, \lb_\sigma \, \l^{\a_{11}} f'(\l\cdot\lb)\right)
}}
Using the properties of the pure spinor measure
\eqn\edLId{\eqalign{
&[d\l]  (\g^m\l)_{r_1}(\g^n\l)_{r_2}(\g^p\l)_{r_3}  (\g_{mnp})_{r_4r_5}
=
\epsilon_{16\,r_1\cdots r_5\g_1\cdots\g_{11}} d\l^{\g_1}\cdots d\l^{\g_{11}}
\cr
&[d\l] \l^{\a} \l^\b \l^{\g}\,\l^\delta = 4\,
\epsilon_{16\,r_1\cdots     r_5\g_1\cdots\g_{11}}     d\l^{\g_1}\cdots
d\l^{\g_{11}}\, \l^{(\a} (T^{-1})^{\b\g\delta)r_1\cdots r_5}
}}
and      the     relation     $\epsilon_{16      r_1\cdots     r_{16}}
\epsilon_{16}^{s_1\cdots s_{16}}= 16!\, \delta^{s_1\cdots s_{16}}_{r_1\cdots r_{16}}$
we get that
\eqn\eStepThree{\eqalign{
\langle 0|C\rangle&=11!5! \int d\l^{\a_1}\wedge\cdots\wedge d\l^{\a_{11}}d\lb_{\a_1}\wedge \cdots
\wedge
d\lb_{\a_{11}}\,e^{-(\l\cdot\lb)\,f(\l\cdot\lb)}\,
f(\l\cdot\lb)^{11}                                                   \,
\cr
&+ 11\, 16!\,4\,    \int       d\l^{\a_1}\wedge\cdots\wedge
d\l^{\a_{11}} d\lb_{\a_1}\wedge \cdots
\wedge
d\lb_{\a_{11}}\,e^{-(\l\cdot\lb)\,f(\l\cdot\lb)}\times
 \lb_\sigma f'(\l\cdot\lb)\, f(\l\cdot\lb)^{10}\cr
&\times     \l^{(\a_{11}}     (T^{-1})^{\b\g\delta)s_1\cdots    s_5}\,
T_{(\a\b\g)r_1\cdots r_5}\,\delta^{\g_1\cdots\g_{11}}_{\a_1\cdots\a_{11}} \delta^{\a_1\cdots\a_{10}\sigma r_1\cdots r_5}_{\g_1\cdots\g_{11} s_1\cdots s_5}
}}
Using that $(T^{-1})^{(\a\b [\g) r_1\cdots r_5]}=0$
we find that $\a_{11}=\sigma$ in the last term, leading to 
\eqn\eStepFour{\eqalign{
\langle 0|C\rangle&= 11!5!\int \prod_{i=1}^{11} d\l^{\a_i}
d\lb_{\a_i}\,e^{-(\l\cdot\lb)\,f(\l\cdot\lb)}\,
f(\l\cdot\lb)^{10}\, (f(\l\cdot\lb)+
 f'(\l\cdot\lb)\,
 (\l\cdot \lb))
}}
Setting $h(\l\cdot\lb)=(\l\cdot\lb)f(\l\cdot\lb)$ this gives
\eqn\eStepFour{\eqalign{
\langle 0|C\rangle&= 11!5!\int \prod_{i=1}^{11} d\l^{\a_i}
d\lb_{\a_i}\,\,(\l\cdot\lb)^{-10}\,e^{-h(\l\cdot\lb)}\,
h(\l\cdot\lb)^{10}\, h'(\l\cdot\lb)
}}
\medskip
\noindent{$\triangleright$} We give another derivation of the same result using some Fierz
identities derived in~\refs{\BerkovitsNG,\MafraJH,\MafraWQ}.

We use the following definition for the normalisations

\eqn\eThreeL{
\int    d^{16}\theta\int     [d\l][d\lb][dr]    \,    e^{-\l\cdot\lb\,
f(\l\cdot\lb)-rM\theta}
\l^\a\l^\b\l^\g f_{\a\b\g}(x,\theta)
= \langle \l^\a\l^\b\l^\g f_{\a\b\g}(x,\theta)\rangle
}
and the Fierz identity established in~\refs{\MafraWQ}
\eqn\eFourLLb{\eqalign{
&\int d^{16}\theta\int [d\l][d\lb][dr] \, e^{-\l\cdot\lb\,f(\l\cdot\lb)-rM\theta}
\l^\a\l^\b\l^\g \lb_\epsilon f_{\a\b\g}^\epsilon(x,\theta)\cr
&
=  {(\l\cdot\lb)\over33}  \,  \left(8  \langle  \l^{(\a}  \l^\b  \l^{\g}
f^{\delta)}_{\a\b\g\delta}\rangle-\langle(\l\g^m)_\epsilon
\g^{(\a\b}_m \l^{\g} \l^{\delta)}
f^{\epsilon}_{\a\b\g\delta}\rangle \right)
}}
The amplitude in~\eStepOne\ takes the form
\eqn\eStepOneBis{\eqalign{
\langle 0|C\rangle&= \int  d^{16}\theta \int [d\l][d\lb][dr]
\,e^{-(\l\cdot\lb)\,f(\l\cdot\lb)}\,f(\l\cdot\lb)^{10} \,(r\theta)^{10} \, \times\cr
&\times 
\left(f(\l\cdot\lb)\,(r\cdot\theta)
(\l\g^m\theta)(\l\g^n\theta)(\l\g^p\theta)(\theta\g_{mnp}\theta)\right.\cr
&\left. +11\, (\lb\cdot\theta) \, (r\cdot\l) f'(\l\cdot\lb) (\l\g^m\theta)(\l\g^n\theta)(\l\g^p\theta)(\theta\g_{mnp}\theta)\right)
}}
The first identity~\eThreeL\ gives
\eqn\eAone{
\langle 0|C\rangle_1 = \langle (r\cdot\theta)^{11} \, f(\l\cdot\lb)^{11} \, (\l\g^m\theta)(\l\g^n\theta)(\l\g^p\theta)(\theta\g_{mnp}\theta)\rangle
}
the second identity~\eFourLLb\ on the second line with 
$f^\epsilon_{\a\b\g\delta}=    \theta^\epsilon    \,    r_\delta    \,
(\g^m\theta)_\a (\g^n\theta)_\b (\g^p\theta)_\g (\theta\g_{mnp}\theta)$
leads to
\eqn\eATwo{\eqalign{
\langle 0|C\rangle_2 &= {2\over 3}\,\langle(r\cdot\theta)^{11} \,
f(\l\cdot\lb)^{10}f'(\l\cdot\lb)                                     \,
(\l\cdot\lb)(\l\g^m\theta)(\l\g^n\theta)(\l\g^p\theta)(\theta\g_{mnp}\theta)\rangle\cr
&-{1\over3} \langle (r\cdot\theta)^{10} 
f(\l\cdot\lb)^{10}f'(\l\cdot\lb)                                     \,
(\l\cdot\lb)(\l\g^s \theta)\, r_\delta \g_s^{(\a\b} \l^\g \l^{\delta)}
\,(\g^m\theta)_\a
(\g^n\theta)_\b(\g^p\theta)_\g(\theta\g_{mnp}\theta)\rangle\cr
&=      {2\over        3}\,\langle(r\cdot\theta)^{11} \,
f(\l\cdot\lb)^{10}f'(\l\cdot\lb)                                     \,
(\l\cdot\lb)(\l\g^m\theta)(\l\g^n\theta)(\l\g^p\theta)(\theta\g_{mnp}\theta)\rangle\cr
&-{1\over6} \langle (r\cdot\theta)^{10} 
f(\l\cdot\lb)^{10}f'(\l\cdot\lb)                                     \,
(\l\cdot\lb)(\l\g^s \theta)\, (r \g_s\g^m\theta)
(\l\g^n\theta)(\l\g^p\theta)(\theta\g_{mnp}\theta)\rangle\cr
}}
where                   we                  used                  that
$(\l\g_s\theta)(\l\g^p\theta)(\theta\g^{mns}\theta)(\theta\g_{mnp}\theta)=0$. 
This expression can be reduced further to 
\eqn\eATwoB{\eqalign{
\langle 0|C\rangle_2 &=      {1\over        2}\,\langle(r\cdot\theta)^{11} \,
f(\l\cdot\lb)^{10}f'(\l\cdot\lb)                                     \,
(\l\cdot\lb)(\l\g^m\theta)(\l\g^n\theta)(\l\g^p\theta)(\theta\g_{mnp}\theta)\rangle\cr
&-{1\over6} \langle (r\cdot\theta)^{10} 
f(\l\cdot\lb)^{10}f'(\l\cdot\lb)                                     \,
(\l\cdot\lb)(\l\g_s \theta)\, (r \g^{sm}\theta)
(\l\g^n\theta)(\l\g^p\theta)(\theta\g_{mnp}\theta)\rangle\cr
}}
Using    the    Fierz    using    that    $3!\,16\,\theta_\a\theta_\b=
(\theta\g_{abc}\theta)(\g^{abc})_{\a\b}$ one shows that
 
\eqn\eTT{
(r\g^{sm}\theta)(\l\g_s\theta)= 4 (r\theta)(\l\g^m\theta)+(\l\g^m\theta)(r\theta)
}
And the total amplitude takes the form
\eqn\eATwoSum{\eqalign{
\langle 0|C\rangle &=  \langle(r\cdot\theta)^{11} \,
f(\l\cdot\lb)^{11}(\l\g^m\theta)(\l\g^n\theta)(\l\g^p\theta)(\theta\g_{mnp}\theta)\rangle\cr
&+  \langle(r\cdot\theta)^{11} \,
f(\l\cdot\lb)^{10}f'(\l\cdot\lb)                                     \,
(\l\cdot\lb)(\l\g^m\theta)(\l\g^n\theta)(\l\g^p\theta)(\theta\g_{mnp}\theta)\rangle\cr
}}
which reproduces~\eStepFour\  after integration  over the $r$  and the
$\theta$ variables.

\listrefs
\bye